\pdfoutput=1

\documentclass[twocolumn]{emulateapj}

\usepackage{amsmath}
\usepackage{graphicx}
\usepackage{color}
\usepackage[colorlinks]{hyperref}
\hypersetup{
    colorlinks,	
    citecolor=blue,
}

\newcommand{\be}{\begin{eqnarray}}
\newcommand{\ee}{\end{eqnarray}}

\shorttitle{Testing reflection models with GRMHD simulations}
\shortauthors{Shashank et al.}

\graphicspath{{./}{figures/}}

\begin{document}

\title{Testing relativistic reflection models with GRMHD simulations\\of accreting black holes}

\author{Swarnim~Shashank\altaffilmark{1}, Shafqat~Riaz\altaffilmark{1}, Askar~B.~Abdikamalov\altaffilmark{1,2,3}, and Cosimo~Bambi\altaffilmark{1,\dag}}

\altaffiltext{1}{Center for Field Theory and Particle Physics and Department of Physics, 
Fudan University, 200438 Shanghai, China. \email[\dag E-mail: ]{bambi@fudan.edu.cn}}
\altaffiltext{2}{Ulugh Beg Astronomical Institute, Tashkent 100052, Uzbekistan}
\altaffiltext{3}{Institute of Fundamental and Applied Research, National Research University TIIAME, Tashkent 100000, Uzbekistan}

\begin{abstract}
X-ray reflection spectroscopy is currently one of the leading techniques for studying the inner part of accretion disks around black holes, measuring black hole spins, and even testing fundamental physics in strong gravitational fields. However, the accuracy of these measurements depends on the reflection models employed for the spectral analysis, which are sometimes questioned. In this work, we use a general relativistic magnetohydrodynamic (GRMHD) code to generate a thin accretion disk in Kerr spacetime and ray-tracing techniques to calculate its relativistically broadened reflection spectrum. We simulate \textsl{NuSTAR} observations and we test the capability of current reflection models to recover the correct input parameters. Our study shows that we can measure the correct input parameters in the case of high inclination angle sources, while we find some minor discrepancies when the inclination angle of the disk is low. 
\end{abstract}


\section{Introduction} \label{sec:intro}

\begin{figure}
    \centering
    \includegraphics[width=0.48\textwidth]{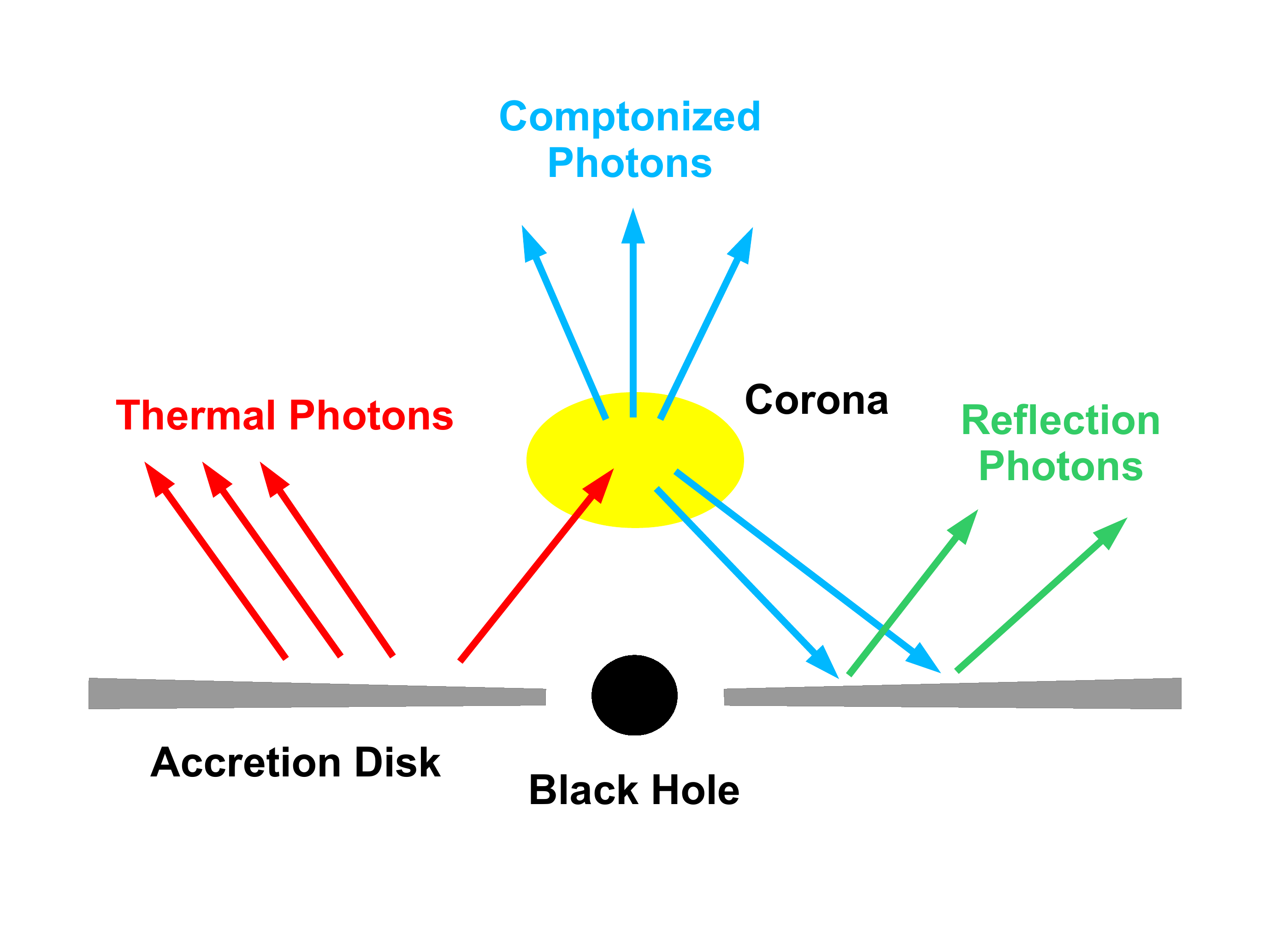}
    \vspace{-1.0cm}
    \caption{Disk-corona model. The reflection spectrum is generated by the illumination of a cold, geometrically thin, and optically thick disk by a hot corona. Figure from~\citet{Bambi:2021chr} under the terms of the Creative Commons Attribution 4.0 International License.}
    \label{fig:disk-corona-model}
\end{figure}

Blurred reflection features are common in the X-ray spectra of accreting black holes \citep[][]{Fabian:1989ej,Tanaka:1995en,Nandra:2007rp,Miller:2009cw}. They are thought to be produced when a hot corona illuminates a cold disk~\citep[][]{Fabian:1995qz}, as illustrated in the cartoon in Fig.~\ref{fig:disk-corona-model} \citep[for a review, see, e.g.,][]{Reynolds:2013qqa,Bambi:2020jpe}. The accretion disk around the black hole is geometrically thin and optically thick. Its thermal spectrum is peaked in the soft X-ray band (0.1-1~keV) in the case of stellar-mass black holes in X-ray binaries and in the UV band (1-100~eV) for supermassive black holes in active galactic nuclei. The ``corona'' consists of some hotter plasma ($\sim 100$~keV) enshrouding the black hole and the inner part of the accretion disk. The corona might be the base of the jet, the atmosphere above the accretion disk, the accretion flow in the plunging region between the inner edge of the disk and the black hole, etc. Thermal photons from the disk can interact with free electrons in the corona through inverse Compton scattering. The spectrum of these Comptonized photons can normally be approximated by a power law with a high-energy cutoff. Such radiation can illuminate the accretion disk. Here we have Compton scattering and absorption followed by fluorescent emission, and the result is a reflection component.

In the rest frame of the material in the accretion disk, the reflection spectrum presents narrow fluorescent emission lines below 10~keV and a Compton hump with a peak normally around 20-30~keV~\cite[][]{Ross:2005dm,Garcia:2010iz}. The most prominent fluorescent emission line is usually the iron K$\alpha$ complex, which is at 6.4~keV for neutral or weakly ionized iron atoms and shifts up to 6.97~keV for hydrogen-like iron ions. The reflection spectrum of the disk as seen by a distant observer appears blurred due to relativistic effects~\cite[][]{Fabian:1989ej,Laor:1991nc,Dauser:2010ne,Bambi:2017khi}. The analysis of these blurred reflection features is potentially a powerful tool for studying the physics and astrophysics in the strong gravity region of accreting black holes.

Thanks to the development of more sophisticated reflection models and new observational facilities, the past 10~years have seen tremendous advancements in the analysis of these relativistic reflection features~\citep[][]{Bambi:2020jpe}. With this technique, today we have the measurement of the spin of about 30 stellar-mass black holes in X-ray binary systems and of about 40 supermassive black holes in active galactic nuclei. It is currently the only mature technique to determine the spins of supermassive black holes, while the spins of stellar-mass black holes can be obtained even from an analysis of the thermal spectrum of the disk~\citep[][]{Zhang:1997dy,McClintock:2013vwa} and from the gravitational wave signal of black hole binaries~\citep[][]{LIGOScientific:2016aoc,Ajith:2009bn}. X-ray reflection spectroscopy is also a powerful technique to test Einstein's theory of General Relativity in the strong field regime and currently provides the most stringent test of the Kerr metric around black holes, somewhat stronger than those inferred from gravitational wave data and significantly more stringent than the constraints inferred from current observations of the shadows of the supermassive black holes in SgrA$^\star$ and M87$^\star$~\citep[][]{Tripathi:2020yts,Tripathi:2021rqs,Zhang:2021ymo}.

However, despite the remarkable progress in the past years, all available relativistic reflection models still rely on a number of assumptions and simplifications~\citep[][]{Bambi:2020jpe}, so caution is needed if we want to use X-ray reflection spectroscopy for precision measurements of accreting black holes. These simplifications are sometimes grouped into four classes: $i)$ simplifications in the description of the accretion disk, $ii)$ simplifications in the description of the hot corona, $iii)$ approximations in the calculations of the reflection spectrum in the rest-frame of the gas, and $iv)$ relativistic effects not properly taken into account. In some cases, specific studies have shown that certain simplifications are acceptable for the current quality of X-ray data and do not introduce any significant bias in the measurement of the properties of the systems. This is the case, for instance, of higher order disk images \citep[see, e.g.,][]{Zhou:2019dfw} and of the radiation emitted from the plunging region \citep[see, e.g.,][]{Cardenas-Avendano:2020xtw}. Some approximations have been removed in more recent reflection models and tested with observations. For instance, there are now models that permit a non-trivial radial disk profile of the ionization parameter and of the electron density \citep{Abdikamalov:2021rty,Abdikamalov:2021ues}, but such improvements do not seem to be strictly necessary for current spin measurements~\citep{Mall:2022llu}. Specific coronal geometries have been investigated in a few studies, but the general conclusion is that phenomenological profiles like a broken power law or a twice broken power law can fit well the current spectra without introducing undesired bias in the parameter estimates \citep[see, e.g.,][]{Wilkins:2011kt,Wilkins:2012zm,Gonzalez:2017gzu,Riaz:2020svt}. Even the returning radiation, namely the radiation that is emitted by the disk and returns to the disk because of strong light bending near the compact object~\citep{Riaz:2020zqb}, has been always neglected in the calculations and only very recently implemented in some models \citep{Dauser:2022zwc}, but it does not seem to affect the estimates of key parameters like the black hole spin.

The standard framework for the description of geometrically thin and optically thick accretion disks around black holes is the Novikov-Thorne disk model~\citep{1973blho.conf..343N,Page:1974he}. All the available relativistic reflection models employ the Novikov-Thorne disk model and thus assume that the velocity field of the disk is Keplerian and that the inner edge of the disk is at the innermost stable circular orbit (ISCO)\footnote{The fact that the inner edge of the disk is at the ISCO radius is an assumption normally employed to measure black hole spins when we fit the data, but all relativistic reflection models have the option to consider even truncated disks, with the inner edge at a radius larger than the ISCO. In what follows, we will use the term Novikov-Thorne disk model to indicate the assumptions of a Keplerian disk and an inner edge at the ISCO radius.}. However, unlike the calculations of the thermal spectrum of the disk, the other ingredients of the Novikov-Thorne model do not enter into the calculations of the reflection spectrum~\citep[see, e.g.,][]{Bambi:2017khi}. Possible deviations from the Keplerian angular velocity of the disk have been tested with observations by \citet{Tripathi:2020wfi}, but they do not seem to be required by the available data.

As a further simplification, current relativistic reflection models normally assume also that the disk is infinitesimally thin. The impact of the thickness of the disk has been studied in a few papers with semi-analytical models, and the conclusion of those studies is that, in general, infinitesimally thin disk models provide a good description in the case of thin disks around fast-rotating black holes, while X-ray reflection spectroscopy measurements can be easily affected by unacceptably large systematic uncertainties in the case of thick disks in sources with very high mass accretion rate \citep{Wu:2007bq,Taylor:2017jep,Riaz:2019bkv,Riaz:2019kat,Abdikamalov:2020oci,Tripathi:2021wap,Jiang:2022sqv}.

In this paper, we want to explore better the accuracy of the Novikov-Thorne disk model in current X-ray reflection spectroscopy measurements, extending existing work in the literature \citep{Reynolds:2007rx,Kinch:2016ipi,Nampalliwar:2022ite}. We use a general relativistic magnetohydrodynamic (GRMHD) code to simulate a geometrically thin accretion disk around a Kerr black hole with a spin parameter of $a_* = 0.98$ and a ray-tracing code to calculate its reflection spectrum in the case of low ($i = 30^\circ$) and high ($i = 70^\circ$) inclination angle of the disk. We simulate a 30~ks observation of a bright Galactic black hole with \textsl{NuSTAR} and we fit the two simulated spectra with reflection models employing infinitesimally thin Novikov-Thorne disks. We find that the fits recover the correct values of the input parameters in the case of a high inclination angle source, while we have some minor discrepancies in the low inclination angle case.

The paper is organized as follows. In Section~\ref{sec:grmhd}, we describes our GRMHD simulation of a thin accretion disk around a black hole. In Section~\ref{sec:raytrace}, we compare the Novikov-Thorne disk and our GRMHD simulated disk and the corresponding iron line profiles produced by the two disk models. In Section~\ref{sec:xspec_simulation}, we simulate two \textsl{NuSTAR} observations of the reflection spectrum of our GRMHD simulated disk and we fit the simulations with relativistic reflection models based on the Novikov-Thorne disk model in order to explore the capabilities of current reflection models to recover the correct input parameters. We discuss our results in Section~\ref{sec:discussion}. Throughout the manuscript, we always use natural units in which $c = G_{\rm N} = 1$, so the gravitational radius is $r_{\rm g} = M$.


\section{GRMHD simulation} \label{sec:grmhd}

We use the {\tt HARMPI} code\footnote{\url{https://github.com/atchekho/harmpi}.} \citep{Gammie:2003rj, Noble:2005gf} in 2.5D to generate a GRMHD simulated thin disk. The initial conditions in the simulation is defined by a Fishbone-Moncrief torus, which is a non-selfgravitating prescription with constant angular momentum per unit inertial mass \citep{FM_torus}. The code itself works in horizon-penetrating Kerr-Schild coordinates but for the analysis it is mapped back to Boyer-Lindquist coordinates. The GRMHD equations \citep{Gammie:2003rj,2008LRR....11....7F} are solved for a Kerr black hole with a source term for cooling the disk, which makes it radiatively efficient  and hence the disk becomes thinner.

\begin{figure}
    \centering
    \includegraphics[width=0.45\textwidth]{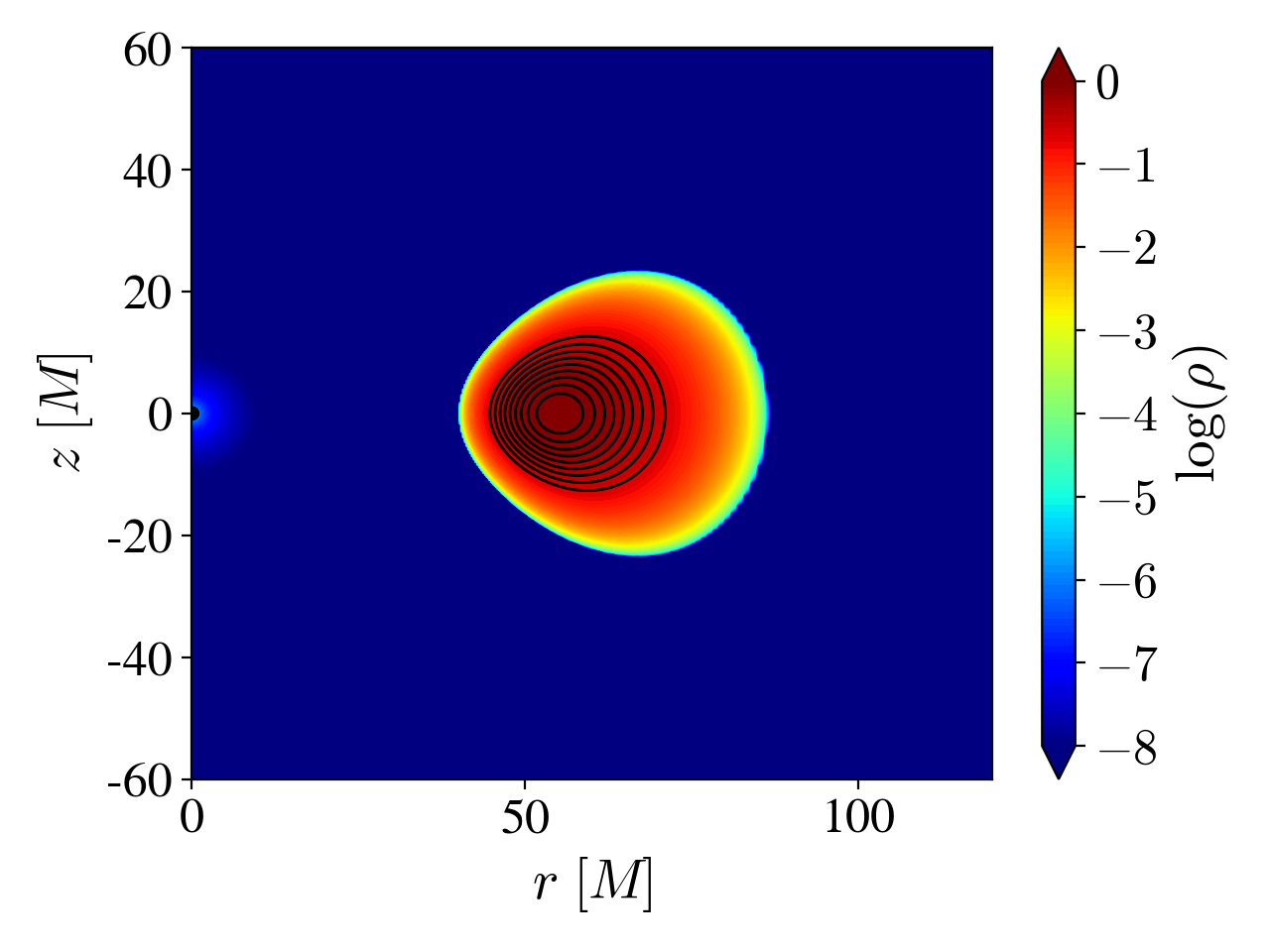}
    \caption{Initial setup of the Fishbone-Moncrief torus. The black lines denote the magnetic fields. Density is normalised.}
    \label{fig:init_torus}
\vspace{1.0cm}
    \centering
    \includegraphics[width=0.45\textwidth]{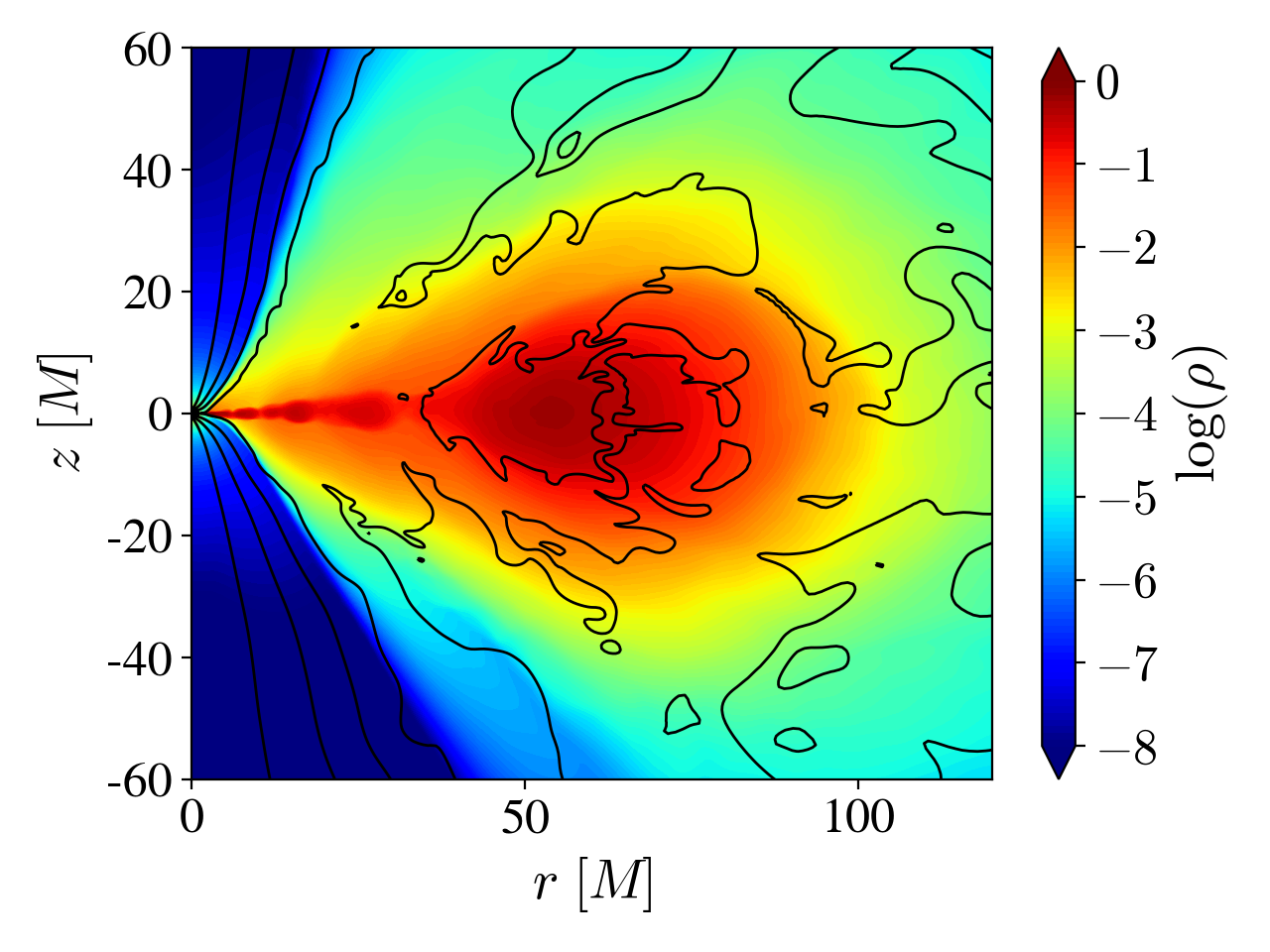}
    \caption{Disk averaged out over time $16000~M$ to $25000~M$. The black lines denote the magnetic fields. Density is normalised.}
    \label{fig:avg_disk}
\end{figure}

The mass conservation equation is
\be
\nabla_{\mu} J^{\mu} = 0 \, ,
\ee
where $J^{\mu} = \rho u^{\mu}$, $\rho$ is rest-mass density of the fluid, and $u^{\mu}$ is its 4-velocity. The equation of the energy-momentum conservation is
\be
\label{eq:energy_conservation}
\nabla_{\mu} T^{\mu}_{\nu} = S_{\nu} \, ,
\ee
where $T^{\mu}_{\nu}$ is the energy-momentum tensor
\be
\hspace{-0.3cm}
T^{\mu}_{\nu} = (\rho + u + P + b^2) u^{\mu} u_{\nu} + \left( P + \frac{b^2}{2} \right) \delta^{\mu}_{\nu} - b^{\mu} b_{\nu} \, ,
\ee
$u$ is the internal energy density, and $P = (\Gamma - 1)u$ is the pressure of an ideal gas. We set $\Gamma = 4/3$, which is the value for a relativistic gas. $b^{\mu}$ is the Lagrangian magnetic 4-field and is related to the laboratory-frame field by $b^{\mu} = B^{\nu} (u^{\mu} u_{\nu} + \delta^{\mu}_{\nu})/{u^t}$.

The source term in Eq.~\ref{eq:energy_conservation} is a radiation 4-force which corresponds to the cooling function $dU/d\tau$ assuming the isotropic co-moving loss of thermal energy \citep{Noble:2008tm, Penna:2010hu}
\be 
S_{\nu} = \left( \frac{dU}{d\tau} \right) u_{\nu} \, .
\ee
Here we use the cooling method of \citet{Penna:2010hu}. The cooling function is used to set a specified value of $H/r$, where $H$ is the half-thickness of the disk at the radial coordinate $r$. The expression of the cooling function is
\be
\frac{dU}{d\tau} = -u~\frac{\log(K/K_c)}{\tau_{\rm cool}}~\exp\left(\frac{-(\theta - \pi/2)^2}{2~\theta_{\rm nocool}^2}\right) \, ,
\ee
where $\tau_{\mathrm{cool}}$ is used to control the cooling rate. In our simulation we use $\tau_{\mathrm{cool}} = 2\pi/\Omega_{\rm K}$, where $\Omega_{\rm K}$ is the Keplerian angular frequency. $K = P/\rho^{\Gamma}$ is the entropy constant of the gas and $K_c$ is the entropy constant of the atmosphere towards which the disk is cooled. $\theta_{\rm nocool}$ is used to provide the target thickness of the disk. We set $\theta_{\rm nocool}$ to $0.1$ for a target $H/r$ of $0.07$ \citep{Penna:2010hu}.

For the initial setup, we use a Fishbone-Moncrief torus with its inner edge at $r_{\mathrm{in}} = 45~M$ and maximum pressure at $r_{\mathrm{max}} = 60~M$. The black hole spin parameter is $a_* = 0.98$. A poloidal magnetic field is seeded to the initial torus with $\beta = 100$, where $\beta = \max P / \max P_b$ and $P_b = b^2 / 2$ is the magnetic pressure. In the code, we impose the following initial 4-vector potential
\be
A_\mu = \left( 0 , 0 , 0 , A_\phi \right) \, ,
\ee
where $A_\phi = {\rm max} \left( \rho/\rho_{\rm max} - 0.2 , 0 \right)$. Fig.~\ref{fig:init_torus} shows the initial setup with magnetic fields.

We run the simulation with a resolution of $768 \times 640 \times 1$ points in polar coordinates. To extract the final accretion disk, we average out the data from time $16000~M$ to $25000~M$, considering time steps of $10~M$. The resulting disk is shown in Fig.~\ref{fig:avg_disk}.


\section{Comparison with the Novikov-Thorne disk model} \label{sec:raytrace}

The Novikov-Thorne disk model \citep{1973blho.conf..343N, Page:1974he} is the standard set-up to describe geometrically thin and optically thick accretion disks around black holes and is implemented in all current relativistic reflection models, which, as a further simplification, normally assume that the disk is infinitesimally thin. Since we want to compare the reflection spectra from a Novikov-Thorne disk and our GRMHD simulated disk, we are interested in two disk properties: the location of the surface of the accretion disk, including the location of the inner edge of the disk, and the velocity of the gas on the surface of the accretion disk. Only these two properties can introduce a difference between the reflection spectra of the two disk models. The other ingredients of the Novikov-Thorne model (e.g., mass accretion rate constant over radius, how the angular momentum is transferred to larger radii and energy is dissipated, etc.) do not enter the calculations of the reflection spectrum.

\begin{figure}
    \centering
    \includegraphics[width=0.48\textwidth]{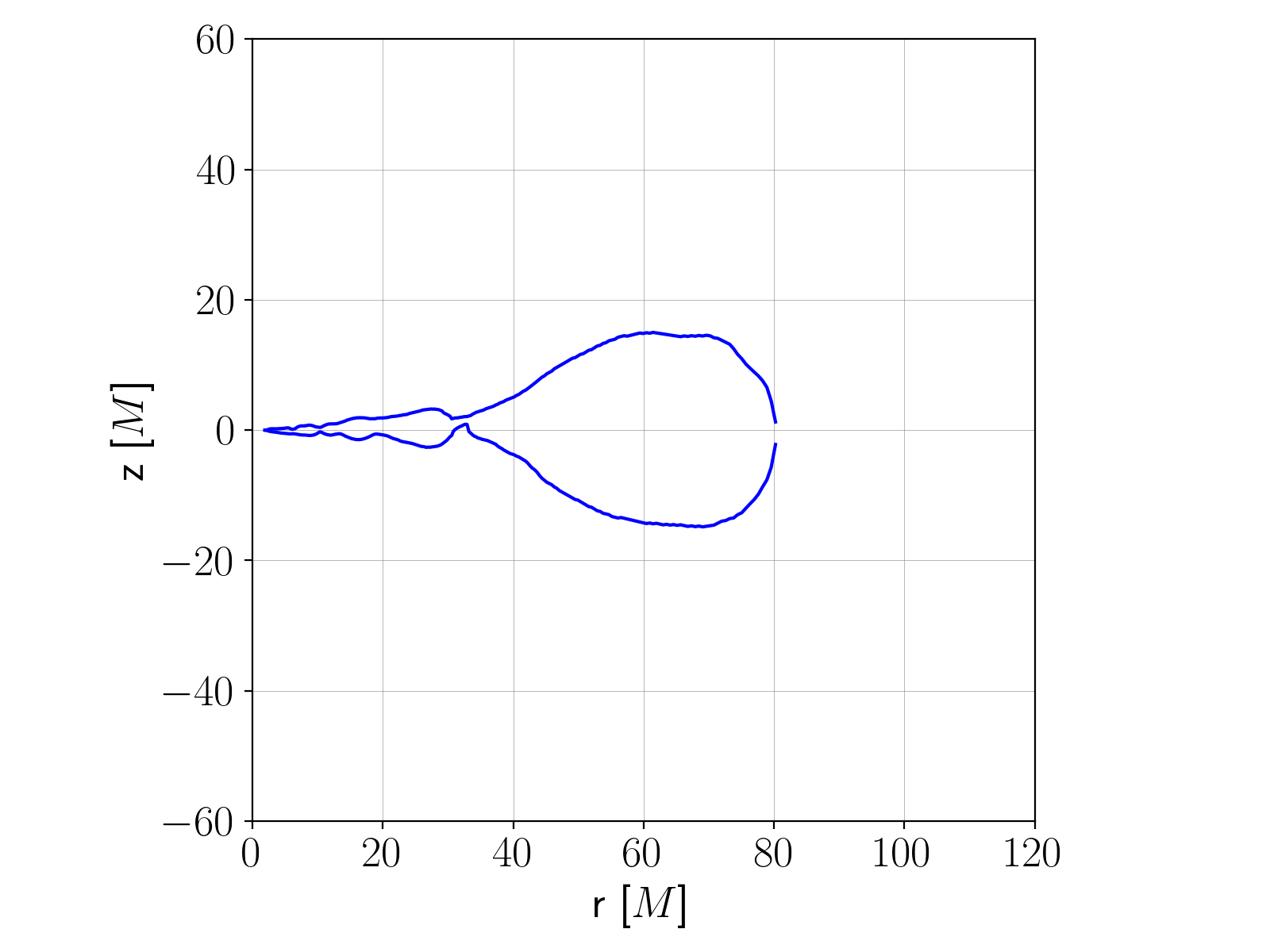}
    \caption{Surface of the accretion disk obtained from the GRMHD simulation and employed for the calculations of the iron line profiles and the reflection spectra.}
    \label{fig:disk_profile}
\vspace{1.0cm}
    \centering
    \includegraphics[width=0.45\textwidth]{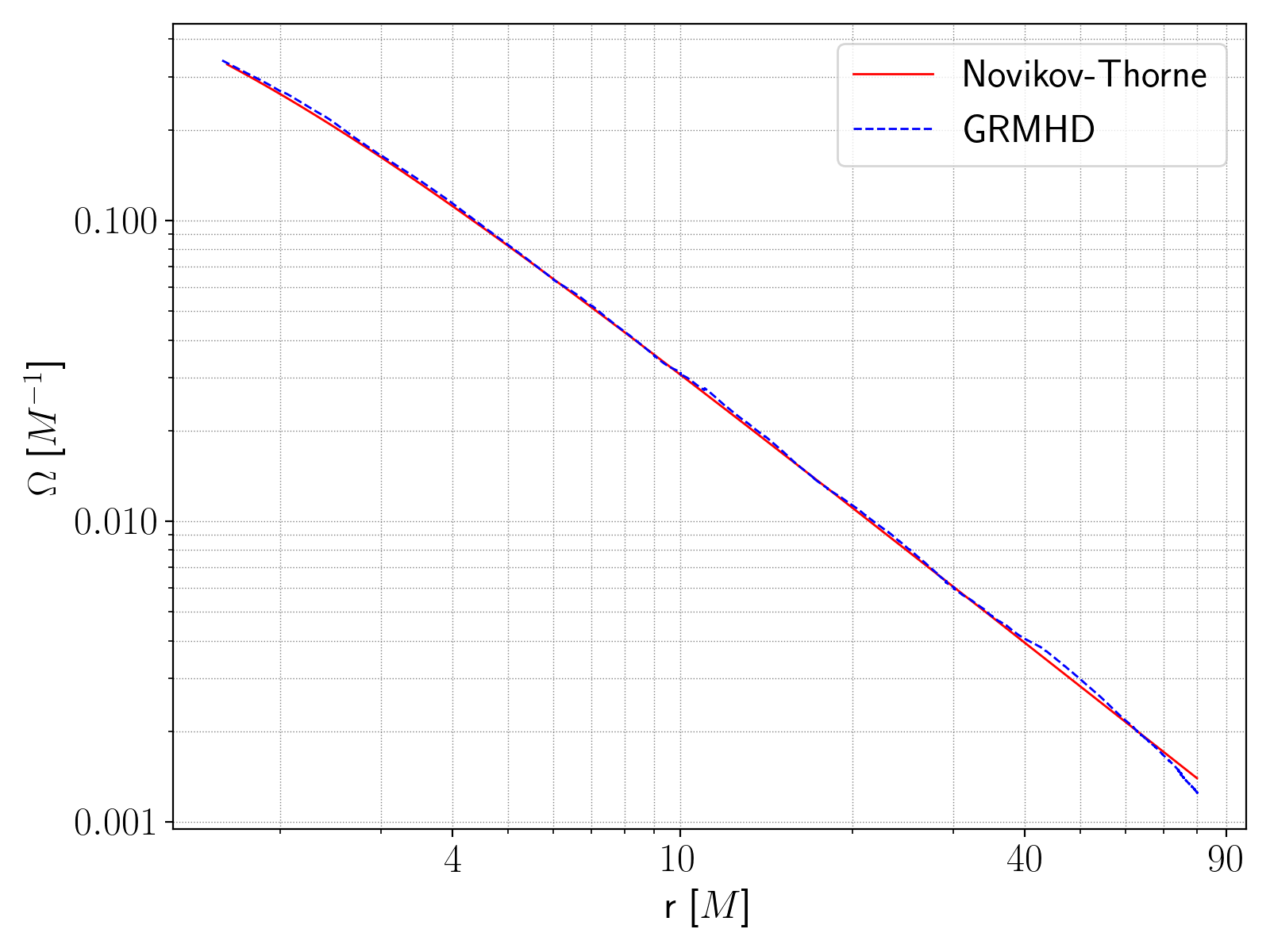}
    \caption{Angular velocities of the gas on the surface of the disk for a Novikov-Thorne accretion disk (solid red curve) and our GRMHD simulated accretion disk (dashed blue curve).}
    \label{fig:angular_vel}
\end{figure}

The surface of the GRMHD simulated disk is determined by using a density cut-off criterion. Choosing $0.1~\rho_{\rm max}$ as the density cut-off, we find that the resulting disk is close to our target thickness in the inner part ($r < 40~M$). The profile of the surface of the accretion disk obtained with such a procedure is shown in Fig.~\ref{fig:disk_profile}. The profile of the angular velocity of the gas on the surface of the accretion disk is plotted in Fig.~\ref{fig:angular_vel} and compared with that in the Novikov-Thorne disk model. The difference between the angular velocities in the two disk models is clearly small at any radial coordinate.

Last, we can calculate the disk spectrum. If we assume that every point on the surface of the disk emits a narrow iron line, the spectrum of the whole disk will be a relativistically broadened iron line. If every point on the surface of the disk emits a reflection spectrum, the spectrum of the whole disk will be a relativistically broadened reflection spectrum. Once we have the spectrum on the surface of the disk, the spectrum of the whole disk as seen by a distant observer can be calculated with well-known ray-tracing techniques, starting from the plane of the distant observer and firing photons to the accretion disk~\cite[see, e.g.,][]{Bambi:2016sac,Bambi:2017khi}. Here we use the ray-tracing code described and tested by \citet{Riaz:2019bkv}. The calculation of the photon trajectory stops when the photon hits the surface of the disk, which is given in Fig.~\ref{fig:disk_profile} for the GRMHD simulated disk and is on the equatorial plane for the Novikov-Thorne disk. When the photon hits the disk, we calculate the redshift factor 
\be
g = \frac{- u_{\rm o}^{\mu} k_{\mu}}{- u_{\rm e}^{\nu} k_{\nu}} \, ,
\ee
where $u_{\rm o}^{\mu} = (1, 0, 0, 0)$ is the 4-velocity of the distant observer, $k^{\mu}$ is the 4-momentum of the photon, and $u_{\rm e}^{\mu}$ is the 4-velocity of the gas on the surface of the disk, which is known in numerical form for the GRMHD simulated disk and we have the analytical expression for the Novikov-Thorne disk. As we have already pointed out, the relevant differences between the two disk models are only the location of the surface of the disk and the velocity of the gas on the surface of the disk. Any other difference between the two disk models does not enter the calculations of the spectrum of the disk.

\begin{figure*}
    \centering
    \includegraphics[width=0.95\textwidth]{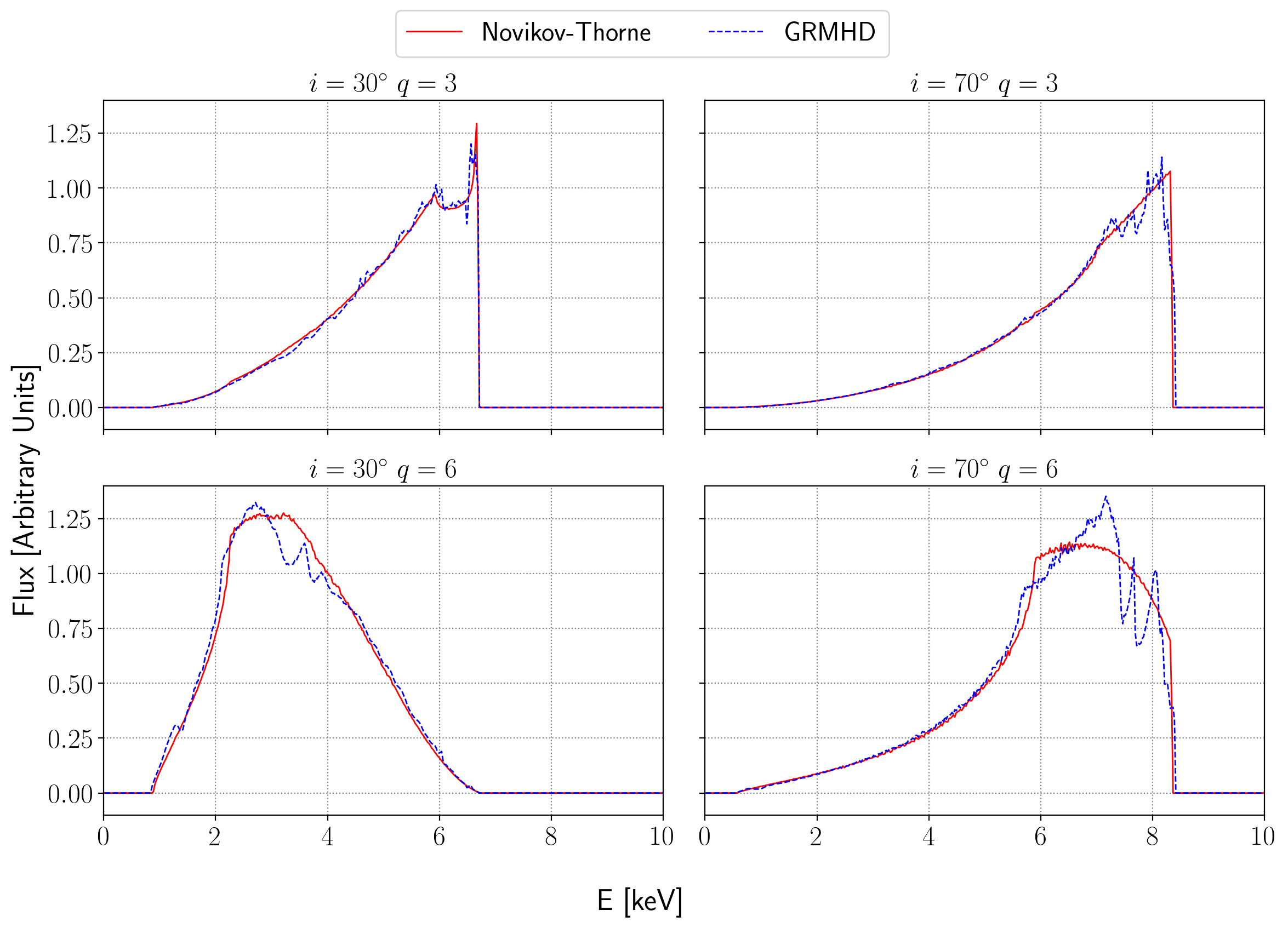}
    \caption{Iron line profiles generated by a Novikov-Thorne accretion disk (solid red curves) and our GRMHD simulated accretion disk (dashed blue curves). The viewing angles of the disks are $i = 30^\circ$ (left panels) and $i = 70^\circ$ (right panels). The emissivity profile of the disks is described by a power law with emissivity indices of $q = 3$ (top panels) and $q = 6$ (bottom panels).}
    \label{fig:ironlines}
\end{figure*}

Assuming that every point of the disk emits a monochromatic line at 6.4~keV in the rest-frame of the gas, the resulting broadened iron lines for the GRMHD and Novikov-Thorne disks are shown in Fig.~\ref{fig:ironlines}. We have considered the case of a low inclination angle of the disk ($i = 30^\circ$) and a high inclination angle of the disk ($i = 70^\circ$), as well as an emissivity profile of the disk described by a power law with emissivity indices of $q=3$ and $q = 6$. As we can see from Fig.~\ref{fig:ironlines}, the difference between the iron line profiles from the two disk models is more pronounced in the case $q = 6$, so we can argue that the GRMHD simulated disk presents more significant deviations from the Novikov-Thorne disk in its inner part.


\begin{figure*}
    \centering
    \includegraphics[width=0.95\textwidth]{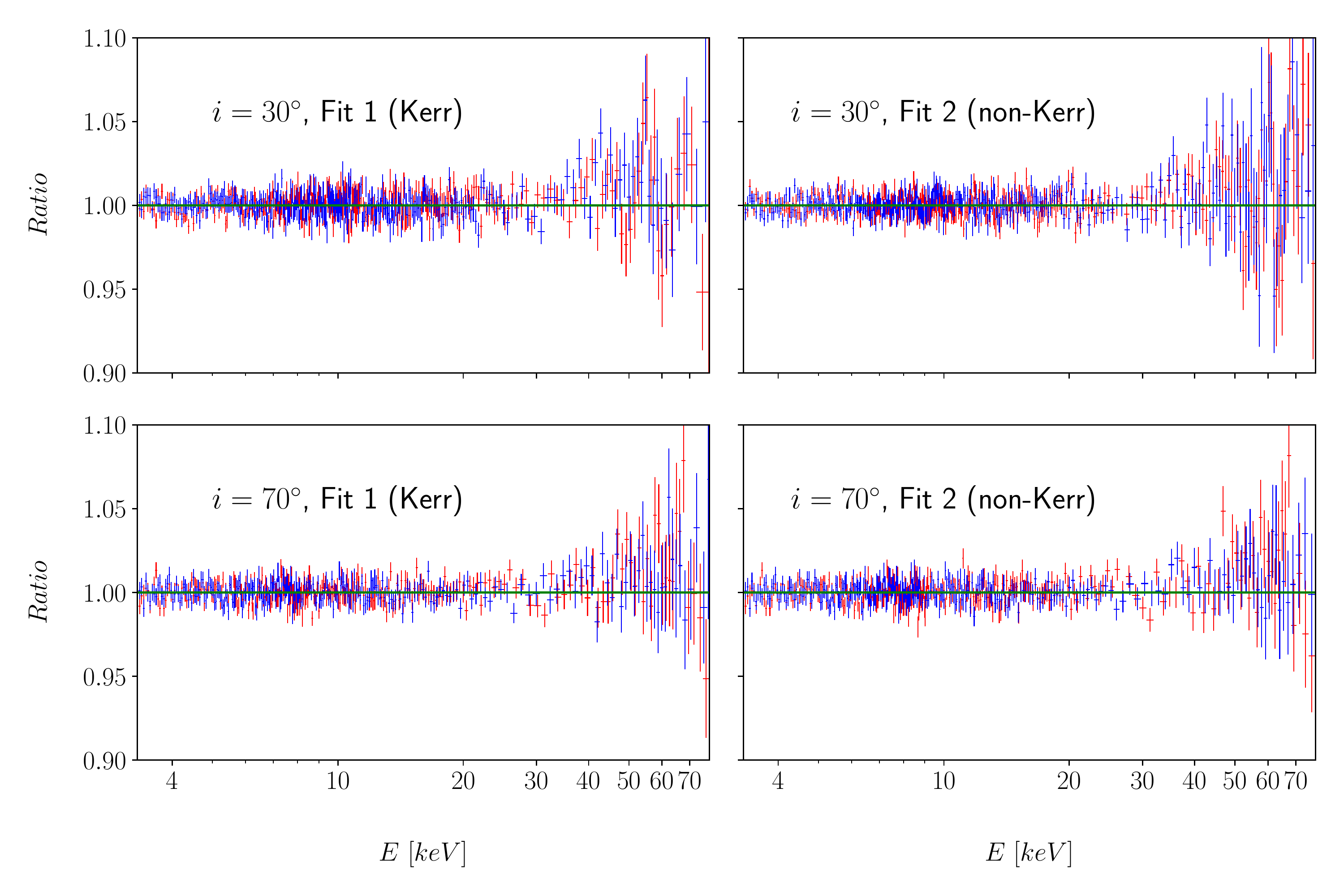}
    \caption{Data to best-fit model ratios for the simulated observations with $i = 30^{\circ}$ and $70^{\circ}$, assuming the Kerr metric ($\alpha_{13} = 0$) or without such an assumption ($\alpha_{13}$ free). Red and blue bars are, respectively, the data from FPMA and FPMB.}
    \label{fig:ratio}
\end{figure*}

\section{Reflection spectra of GRMHD simulated disks} \label{sec:xspec_simulation}

Employing the ray-tracing code discussed in the previous section and assuming that every point of the disk emits a reflection spectrum, we can calculate the relativistically blurred reflection spectrum of the disk as see by a distant observer. For the reflection spectrum in the rest-frame of the gas, we use the table of {\tt xillver}~\citep{Garcia:2010iz,Garcia:2013lxa}. Now we consider only the case with emissivity index of $q=6$, as we want to study the differences between the reflection spectra from the GRMHD simulated disk and the Novikov-Thorne disk. As seen in the previous section for an iron line, for $q=3$ the differences between the two spectra are smaller. We still calculate two cases, one with a low inclination angle of the disk ($i = 30^\circ$) and the other one with a high inclination angle ($i = 70^\circ$).

With the two theoretical spectra for $i = 30^\circ$ and $70^\circ$, we simulate the observation of a bright Galactic black hole with \textsl{NuSTAR}~\citep[][]{NuSTAR:2013yza}. \textsl{NuSTAR} is the most suitable X-ray observatory currently available for X-ray reflection spectroscopy measurements of bright sources, as its detectors are normally not affected by pile-up and, thanks to their broad energy band, we can simultaneously observe the broadened iron line and the Compton hump of the reflection spectrum of the source. To produce the simulated observations, we use the \texttt{fakeit} command in XSPEC~\citep[][]{xspec} together with the background, ancillary, and response files (for both Focal Plane Module A and B) of EXO~1846--031 obtained in the analysis of~\citet{Riaz:2022rlx}.

\begin{table*}
\centering
\caption{\rm Input parameters and the best-fit values for simulations. Fit~1 refers the best-fit model when we assume the Kerr spacetime ($\alpha_{13}=0$ in {\tt relxill\_nk}) and Fit~2 refers the best-fit model when we relax this hypothesis ($\alpha_{13}$ is free in {\tt relxill\_nk}). The reported uncertainties correspond to 90\% confidence level for one relevant parameter. $^\star$ indicates that the parameter is frozen in the fit. If the upper/lower uncertainty is not  reported, it means that the 90\% confidence level limit is not within the boundary of the parameter. \label{t-fits}}
{\renewcommand{\arraystretch}{1.3}
\begin{tabular}{lccc|ccc}
\hline\hline
& \multicolumn{3}{c}{Spectrum~1} & \multicolumn{3}{c}{Spectrum~2} \\
& \hspace{0.5cm} Input \hspace{0.5cm} & \hspace{0.5cm} Fit~1 \hspace{0.5cm} & \hspace{0.5cm} Fit~2 \hspace{0.5cm} & \hspace{0.5cm} Input \hspace{0.5cm} & \hspace{0.5cm} Fit~1 \hspace{0.5cm} & \hspace{0.5cm} Fit~2 \hspace{0.5cm} \\
\hline
\texttt{tbabs} &&&&&& \\
$N_{\rm H} / 10^{20}$ cm$^{-2}$ & $6.74$ & $6.74^\star$ & $6.74^\star$ & $6.74$ & $6.74^\star$ & $6.74^\star$ \\
\hline
\texttt{cutoffpl} &&&&&& \\
$\Gamma$ & $1.7$ & $1.64_{-0.03}^{+0.04}$ & $1.64_{-0.03}^{+0.03}$ & $1.7$ & $1.691_{-0.009}^{+0.009}$ & $1.690_{-0.009}^{+0.009}$ \\
$E_{\rm cut}$ [keV] & $300$ & $283_{-22}^{+33}$ & $265_{-40}^{+24}$ & $300$ & $289_{-10}^{+13}$ & $289_{-10}^{+13}$ \\
\hline
\texttt{relxill\_nk} &&&&&& \\
$q$ & $6$ & $5.9_{-0.3}^{+0.5}$ & $5.058_{-0.150}^{+0.010}$ & $6$ & $5.7_{-0.7}^{+0.8}$ & $5.0_{-0.9}^{+1.3}$ \\
$a_*$ & $0.98$ & $0.993_{-0.006}^{}$ & $0.998_{-0.011}^{}$ & $0.98$ & $0.973_{-0.009}^{+0.005}$ & $0.964_{-0.015}^{+0.017}$ \\
$\alpha_{13}$ & $0$ & $0^\star$ & $-0.4_{}^{+0.4}$ & $0$ & $0^\star$ & $-0.32_{-0.14}^{+0.39}$ \\
$i$ [deg] & $30$ & $30_{-3}^{+7}$ & $23.68_{-0.22}^{+3.14}$ & $70$ & $67.8_{-1.8}^{+1.6}$ & $68.5_{-3.0}^{+1.3}$ \\
$\log\xi$ [erg~cm~s$^{-1}$] & $3.1$ & $3.19_{-0.05}^{+0.07}$ & $3.19_{-0.05}^{+0.06}$ & $3.1$ & $3.14_{-0.03}^{+0.04}$ & $3.14_{-0.03}^{+0.04}$ \\
$A_{\rm Fe}$ & $1$ & $1.53_{-0.33}^{+0.24}$ & $1.490_{-0.018}^{+0.150}$ & $1$ & $0.99_{-0.04}^{+0.07}$ & $0.98_{-0.04}^{+0.07}$ \\
\hline
$\chi^2/\nu$ && $3251.94/3301$ & $3251.72/3300$ && $3588.06/3428$ & $3587.27/3427$ \\
&& $=0.98513$ & $=0.98536$ && $=1.04669$ & $=1.04676$ \\
\hline\hline
\end{tabular}}
\end{table*}

In XSPEC language, the model of the simulated observations reads
\begin{align*}
{\tt tbabs\times(cutoffpl + reflection)}.   
\end{align*}
\texttt{tbabs}~\citep[][]{Wilms:2000ez} takes the Galactic absorption into account and has one parameter, the hydrogen column density $N_{\rm H}$, which we set to $6.74\times10^{22}~\mathrm{cm}^{-2}$. \texttt{cutoffpl} models the power-law continuum from the corona: we set the photon index to $\Gamma = 1.7$ and the high-energy cutoff to $E_{\rm cut} = 300$~keV. \texttt{reflection} is the theoretical reflection spectrum calculated by the ray-tracing code with the GRMHD simulated disk. In the GRMHD simulation, the black hole spin is $a_* = 0.98$. The ray-tracing code assumes that the inclination angle of the disk is either $i = 30^\circ$ or $70^\circ$ and that the emissivity profile of the disk is described by a power law with an emissivity index of $q=6$. For the reflection spectrum in the rest-frame of the gas extracted from the {\tt xillver} table, we choose the ionization parameter $\log\xi = 3.1$ ($\xi$ in units of erg~cm~s$^{-1}$) and the disk’s iron abundance $A_{\rm Fe} = 1$ (Solar abundance). Since we want to simulate a bright Galactic black hole, we impose that the photon flux in the energy band 1 to 10~keV is approximately $1\times10^{-8}$~erg~cm$^{-2}$~s$^{-1}$. We adjust the normalization of the power-law component and the reflection component so that both components contribute equally to the total photon flux. We set the exposure time to 30~ks. With these choices, we have about 7~million photons in the energy range 3 to 78~keV. We bin the spectra using the \texttt{grppha} routine to have at least 30 counts per energy bin.

We fit the two simulated spectra for $i = 30^\circ$ and $70^\circ$ with the following model
\begin{align*}
\label{fit-model}
    {\tt tbabs\times(cutoffpl + relxill\_nk)}.
\end{align*}
\texttt{relxill\_nk}~\citep{Bambi:2016sac,Abdikamalov:2019yrr} is an extension to non-Kerr spacetimes of the relativistic reflection model \texttt{relxill}~\citep[][]{Dauser:2013xv, Garcia:2013lxa}. With respect to \texttt{relxill}, we have one more parameter, which is called the deformation parameter and quantify possible deviations from the Kerr solution. Here we use the version of \texttt{relxill\_nk} in which the deformation parameter is the parameter $\alpha_{13}$ of the Johannsen metric~\citep[][]{Johannsen:2013szh}. If we set $\alpha_{13} = 0$, we recover the Kerr metric and \texttt{relxill\_nk} formally reduces to \texttt{relxill}. In our fits, we always assume that the inner edge of the accretion disk is at the ISCO, which, when written in units of $r_{\rm g}$, is only determined by the black hole spin parameter $a_*$ (in the Kerr spacetime) and by $a_*$ and $\alpha_{13}$ (when $\alpha_{13}$ is free in the fit).

\begin{figure}[h]
    \centering
    \includegraphics[width=0.48\textwidth]{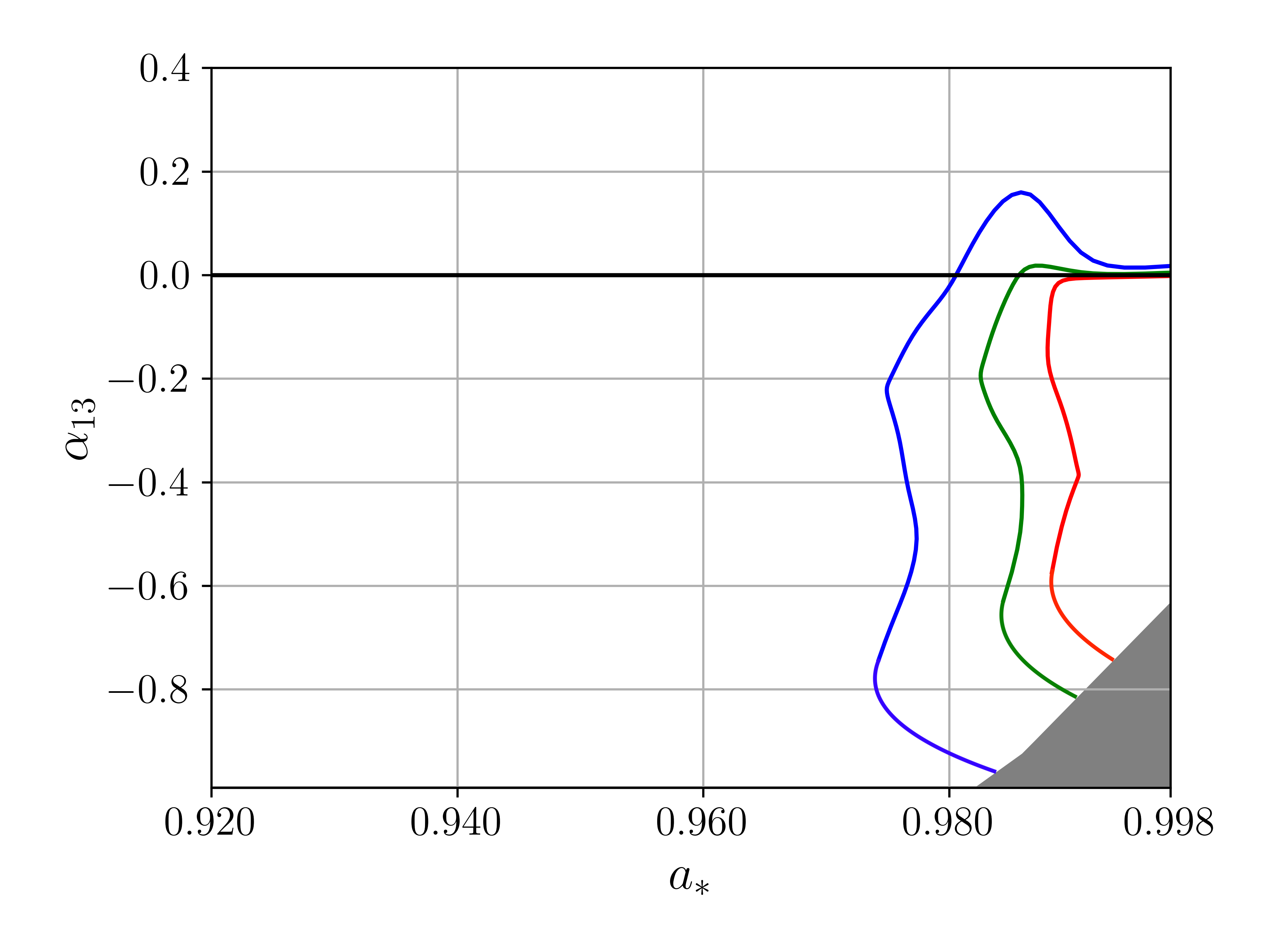}
    \vspace{-0.8cm}
    \caption{Constraints on the spin parameter $a_*$ and the Johannsen deformation parameter $\alpha_{13}$ from the simulated observation with $i = 30^{\circ}$. The red, green, and blue curves are, respectively, the 68\%, 90\%, and 99\% confidence level contours for two relevant parameters. The grey region is excluded in our analysis because it is the parameter space with pathological spacetimes.}
    \label{fig:contPlot30deg}
\vspace{1.0cm}
    \centering
    \includegraphics[width=0.48\textwidth]{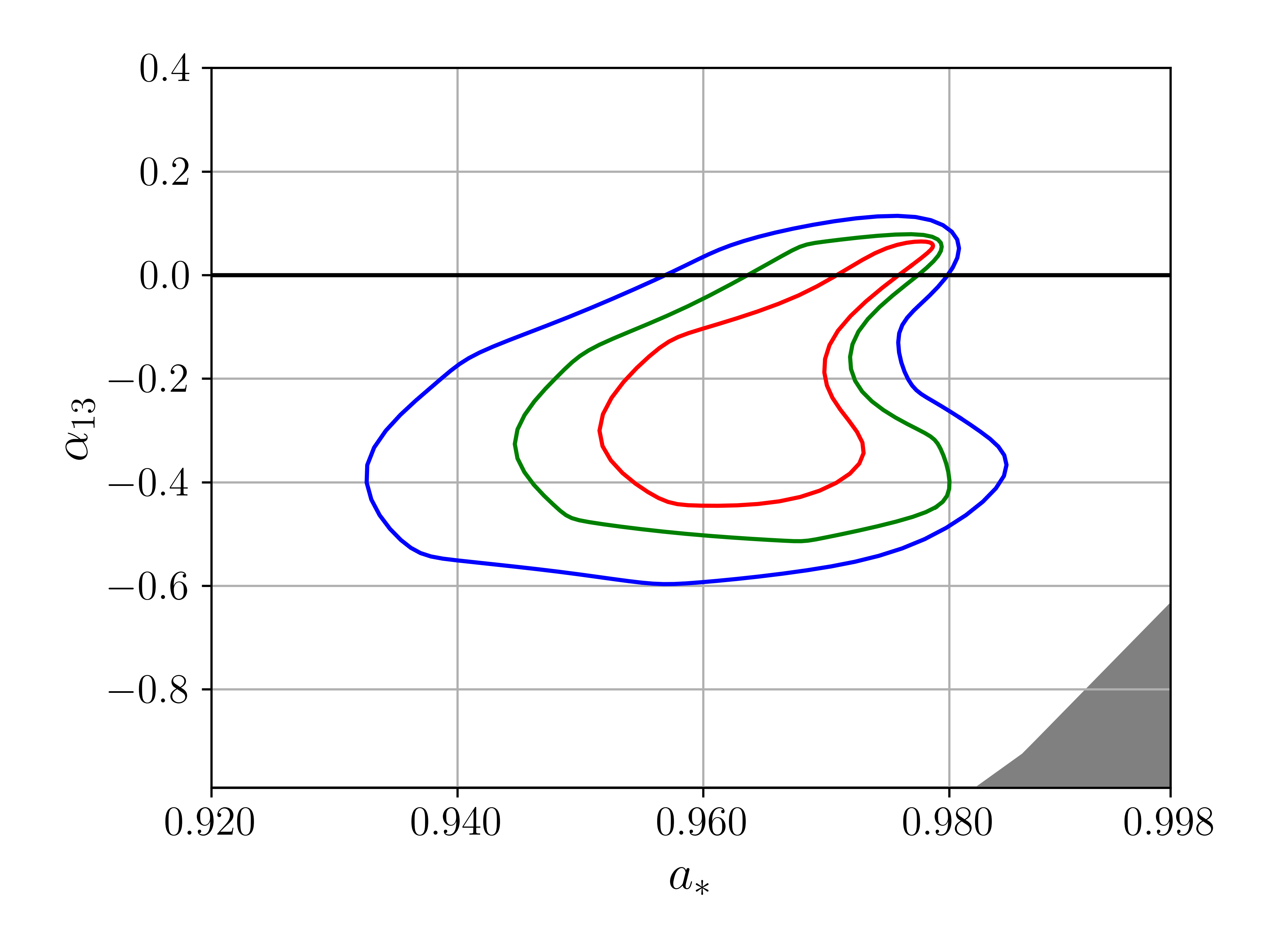}
    \vspace{-0.8cm}
    \caption{As in Fig.~\ref{fig:contPlot30deg} for the simulation with $i = 70^{\circ}$.}
    \label{fig:contPlot70deg}
\end{figure}

First, we fit the two simulated spectra assuming the Kerr metric. The results of our fits are shown in Tab.~\ref{t-fits} (column Fit~1, where $\alpha_{13} = 0$) and in Fig.~\ref{fig:ratio} (left panels). We note that we have also fitted the spectra with the normal {\tt relxill} and with {\tt relconv$\times$xillver}, obtaining consistent results. The fit with {\tt relconv$\times$xillver} for the spectrum with $i = 30^\circ$ presents some minor discrepancies, but this is still consistent with the fact that {\tt relconv$\times$xillver} is an angle-averaged model \citep[see, e.g.,][]{Tripathi:2020cje}.

We repeat the fits with a free deformation parameter $\alpha_{13}$ to check the capability of our model to test the Kerr metric~\citep{Bambi:2015kza}. The results of these new fits are shown in Tab.~\ref{t-fits} (column Fit~2) and in Fig.~\ref{fig:ratio} (right panels). We use the \texttt{steppar} command in XSPEC to obtain the constraints on the spin parameter $a_*$ and Johannsen deformation parameter $\alpha_{13}$. The results are shown in Fig.~\ref{fig:contPlot30deg} and Fig.~\ref{fig:contPlot70deg}, respectively, for the simulations with $i = 30^\circ$ and $70^\circ$.


\begin{figure*}[t]
    \centering
    \includegraphics[width=0.45\textwidth]{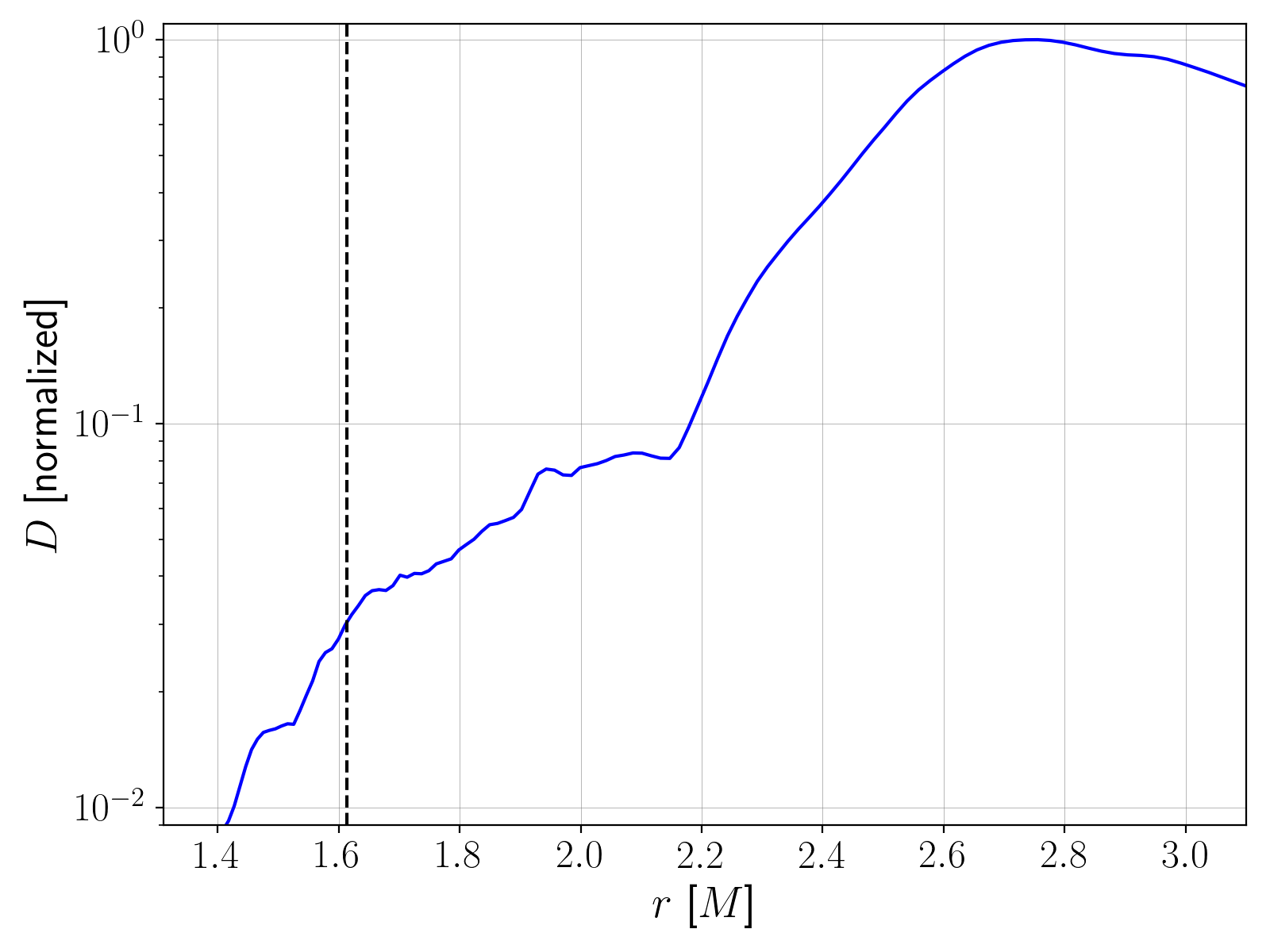}
    \hspace{0.5cm}
    \includegraphics[width=0.45\textwidth]{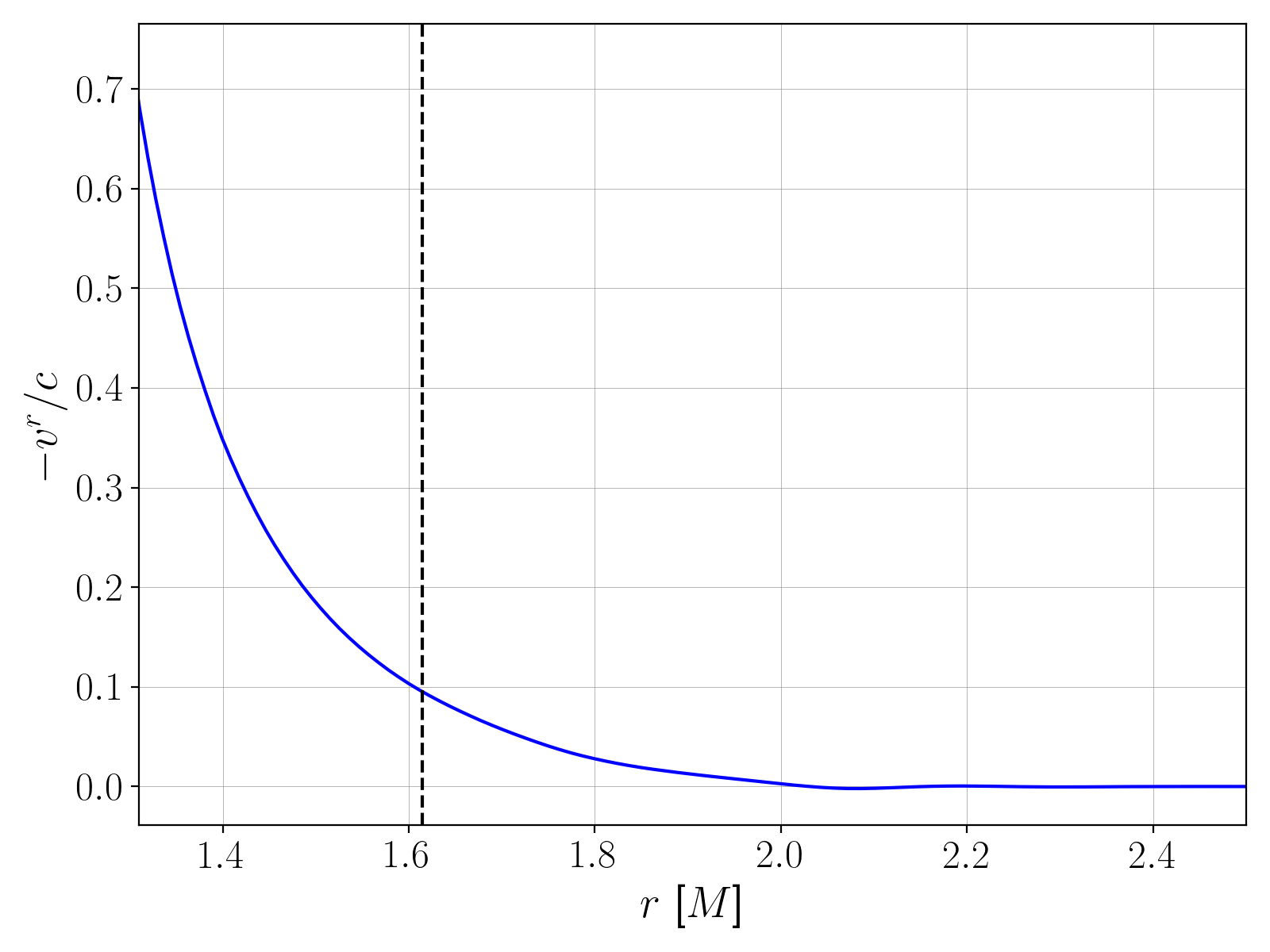}
    \caption{Radial profiles of the density $D = \rho W$ (left panel) and radial 3-velocity $v^r = u^r/W$ (right panel) on the equatorial plane at small radii. The vertical dashed line indicates the radial coordinate of the ISCO radius for a Kerr black hole with spin parameter $a_* = 0.98$.}
    \label{fig:dvr}
\end{figure*}

\section{Discussion and conclusions}\label{sec:discussion}

From Tab.~\ref{t-fits} and Fig.~\ref{fig:ratio}, we can immediately conclude that the relativistic reflection model employing Novikov-Thorne disks fits the simulated data well. In Fig.~\ref{fig:ratio}, we do not see unresolved features. In Tab.~\ref{t-fits}, the estimates of the model parameters match well with their input values, even if we have to note some difference between the low inclination case (Spectrum~1) and the high inclination one (Spectrum~2). In the latter case, we indeed recover the correct input values, and this is true in Fit~1 in which we assume the Kerr metric as well as in Fit~2 where the deformation parameter $\alpha_{13}$ is left free in the fit. In the simulation with a low inclination angle disk, we see instead some minor discrepancy between the input value and the best-fit value of some parameters. The spin parameter $a_*$, the ionization parameter $\xi$, and the iron abundance $A_{\rm Fe}$ are slightly overestimated, independent of the choice to have $\alpha_{13}$ frozen to zero or free in the fit. In the fit with free $\alpha_{13}$, we also find that the value of the emissivity profile is underestimated. We note that such a trend does not depend on the specific simulation: even repeating the simulations and the fits, we find that the fit recovers well the input values of the high inclination angle case and there are minor discrepancies for the simulation with $i = 30^{\circ}$.

In our simulations, we have considered a 30~ks observation with \textsl{NuSTAR} of a putative very bright Galactic black hole, resulting in about 7~million photons in the detector energy band. The quality of the two simulated spectra is quite good and the estimate of the black hole spin is precise, as we can see from Tab.~\ref{t-fits}. Typical \textsl{NuSTAR} observations of Galactic black holes do not have such a high photon count and presumably we would not be able to see the minor discrepancy in the estimate of some parameters in the case of the low inclination angle spectrum. In the end, our results suggest that infinitesimally thin Keplerian disks with inner edge at the ISCO can describe well real disks around black holes.

The validity of the Novikov-Thorne disk model was tested with GRMHD simulations in \citet{Noble:2010mm} and \citet{Penna:2010hu}, finding different conclusions. \citet{Noble:2010mm} employs a model with a highly magnetized corona and find that their GRMHD simulated disk presents important deviations from the Novikov-Thorne disk model. The simulations in \citet{Penna:2010hu} do not include any corona (as it is plausible to assume in the high/soft state of a black hole binary) and their initial magnetic field consists of multiple poloidal field loops. The authors do not find significant differences between their GRMHD simulated disk and the Novikov-Thorne disk model. The presence of a corona and the initial conditions of the magnetic field appear thus crucial to determine the behavior of the accretion flow around the black hole and, in turn, the validity of the Novikov-Thorne model. Since our initial setup is similar to that employed in \citet{Penna:2010hu}, it is understandable that we find that a relativistic reflection model employing Novikov-Thorne disks fits well the reflection spectra of our GRMHD simulated disk. For a different initial setup, such a conclusion is not guaranteed. Since we do not have yet a good understanding of the magnetic fields around accreting black holes, we cannot say which GRMHD simulation describes better a real source.

Iron line profiles from MHD/GRMHD simulated disk have been reported in \citet{Reynolds:2007rx}, \citet{Kinch:2016ipi}, and \citet{Nampalliwar:2022ite}, while we are not aware of any work in which it is calculated the full reflection spectrum except the present study. \citet{Kinch:2016ipi} and \citet{Nampalliwar:2022ite} do not report any quantitative analysis to compare the iron line profiles from their GRMHD disks and those predicted from Novikov-Thorne disks, but from the figures in those papers we can conclude that the iron lines reported in \citet{Kinch:2016ipi} are typically quite different from those predicted by the Novikov-Thorne model while the iron line reported in \citet{Nampalliwar:2022ite} present small deviations from the iron line of Novikov-Thorne disks. Even in this case, the choice of the initial setup seems to be the key-point to determine the final result. We note that the setup employed in \citet{Nampalliwar:2022ite} is similar to ours in the present work and their results and conclusions appear to match well with ours.

\citet{Reynolds:2007rx} is the only work already present in the literature reporting a quantitative and systematic study of the capability of iron line models based on Novikov-Thorne disks to recover the correct values of the black hole spins. \citet{Reynolds:2007rx} use high-resolution 3D MHD simulations in a pseudo-Newtonian potential. They find that black hole spins are slightly overestimated, which follows from the fact that in their simulations the inner edge of the disk is not at the ISCO but it extends to slightly smaller radii, mimicking a lower value of the ISCO radius and therefore a higher black hole spin. They find that the systematic error in the measurement of the black hole spin parameter decreases as the actual black hole spin increases.

In our study, the reflection surface of the disk is defined by a cutoff density. This is quite a simple criterion and has the advantage that does not require to fix the scale of the system. \citet{Reynolds:2007rx} and \citet{Nampalliwar:2022ite} define the reflection surface as the surface with optical depth $\tau = 1$. For example, the plunging region can be optically thick even if the gas density is low. However, as discussed in \citet{Reynolds:2007rx}, the ``reflection edge'' can still be around the inner edge of the accretion disk: as the gas density drops dramatically in the plunging region, the gas becomes highly ionized. In such a case, the interactions between the photons from the corona and the material in the plunging region are dominated by Compton scattering. The resulting reflection spectrum has no emission lines and has instead the same shape as the incident continuum. In the analysis of the spectrum of the source, the reflection spectrum from the plunging region contributes to the continuum, without affecting the analysis of the reflection features of the disk and, in turn, the estimate of the parameters of the model.

Fig.~\ref{fig:dvr} shows the radial profile of the gas density in the rest-frame of the gas $D$ and the gas radial 3-velocity $v^r$ on the equatorial plane near the black hole from our GRMHD simulation
\be 
D= \rho W \, , \quad v^r = u^r/W \, ,
\ee
where $\rho$ is the rest-mass density of the fluid, $W = \alpha u^t$, $\alpha$ is the lapse function, and $u^\mu$ is the 4-velocity of the fluid \citep[see, e.g.,][]{2008LRR....11....7F}. The vertical dashed lines in Fig.~\ref{fig:dvr} mark the radial coordinate of the ISCO for a Kerr black hole with a spin parameter of $a_* = 0.98$. From the GRMHD simulation we thus recover that the gas density decreases dramatically in the plunging region and the radial velocity of the gas increases. This is consistent with the results reported by \citet{Reynolds:2007rx} and is crucial for the validity of reflection based black hole spin measurements. Such a quick decrease of the density in the plunging region indeed determines the reflection edge of the disk and justifies the assumption of current reflection based black hole spin measurements of no reflection emission inside the ISCO.

In conclusion, here we have presented a study to understand the capability of current relativistic reflection models based on Novikov-Thorne disks to measure the properties of accreting black holes. With a GRMHD code, we have generated a thin accretion disk around a Kerr black hole with a spin parameter of $a_* = 0.98$ and then we have simulated two \textsl{NuSTAR} observations of a bright Galactic black hole, one for a low inclination angle of the disk and the other one for a high value of the inclination angle. In our spectral analysis of these two simulated observations we are able to recover the correct input parameters. Our results appear to be consistent with previous studies already present in the literature and based only on iron line profiles \citep{Reynolds:2007rx,Nampalliwar:2022ite}. We note that the choice of the initial conditions of the GRMHD simulation, in particular the choices of the initial magnetic field and the possible presence of a highly magnetized corona, seem to play a fundamental role in the final results.


\begin{acknowledgments}
This work was supported by National Natural Science Foundation of China (NSFC), Grant No. 11973019, the Natural Science Foundation of Shanghai, Grant No. 22ZR1403400, the Shanghai Municipal Education Commission, Grant No. 2019-01-07-00-07-E00035, and Fudan University, Grant No. JIH1512604.
S.S. also acknowledges support from the China Scholarship Council (CSC), Grant No.~2020GXZ016646.
\end{acknowledgments}

\bibliographystyle{apj}
\bibliography{references}

\end{document}